# Coupled Phonons, Magnetic Excitations and Ferroelectricity in AlFeO$_3$: Raman and First-principles Studies

Pradeep Kumar[1], Achintya Bera[1], D. V. S. Muthu[1], Sharmila N. Shirodkar[2], Rana Saha[3], Ajmala Shireen[3], A. Sundaresan[3], U. V. Waghmare[2], A. K. Sood[1,*] and C. N. R. Rao[3]

[1]Department of Physics, Indian Institute of Science, Bangalore -560012, India

[2]Theoretical Sciences Unit, Jawaharlal Nehru Centre for Advanced Scientific Research, Jakkur, P.O. Bangalore -560064, India

[3] Chemistry and Physics of Materials Unit, New Chemistry Unit and International Centre for Materials Science, Jawaharlal Nehru Centre for Advanced Scientific Research, Jakkur, P.O. Bangalore-560064, India

## ABSTRACT

We determine the nature of coupled phonons and magnetic excitations in AlFeO$_3$ using inelastic light scattering from 5 K to 315 K covering a spectral range from 100-2200 cm$^{-1}$ and complementary first-principles density functional theory-based calculations. A strong spin-phonon coupling and magnetic ordering induced phonon renormalization are evident in (a) anomalous temperature dependence of many modes with frequencies below 850 cm$^{-1}$, particularly near the magnetic transition temperature $T_c$ ~ 250 K, (b) distinct changes in band positions of high frequency Raman bands between 1100-1800 cm$^{-1}$, in particular a broad mode near 1250 cm$^{-1}$ appears only below $T_c$ attributed to the two-magnon Raman scattering. We also observe weak anomalies in the mode frequencies at ~ 100 K, due to a magnetically driven ferroelectric phase transition. Understanding of these experimental observations has been possible on the basis of first-principles calculations of phonons spectrum and their coupling with spins.

*Corresponding author: email: asood@physics.iisc.ernet.in, Ph: +91-80-22932964

## 1. INTRODUCTION

Materials that exhibit co-occurrence of both magnetic and ferroelectric order parameters have generated enormous interest in recent years because of fundamental issues related to the coupling between spin, orbital, charge and lattice degrees of freedom as well as for their potential applications [1-5]. For applications, it is desirable to have materials with magneto-electric properties around room temperature, which is often not realized in many magneto-electric materials in which magnetic ordering is the primary driving force. In this context, $AlFeO_3$ exhibiting ferrimagnetism and possible magneto-electric coupling is very promising with a paramagnetic to ferrimagnetic transition temperature $T_c \sim$ 250 K [6-7]. Another attractive feature is its environment friendly nature as compared to other lead based multiferroics.

In $AlFeO_3$, cations occupy four distinct crystallographic sites: cations $Fe_1$, $Fe_2$ and $Al_2$ are octahedrally coordinated by oxygen whereas $Al_1$ is tetrahedrally coordinated. Structural analysis of $AlFeO_3$ [7] shows significant distortion of the $FeO_6$ octahedra, while oxygen tetrahedron around $Al_1$ is quite regular. The cause for the local deformation of lattice has been attributed to the difference between octahedral radii of $Fe^{3+}$ and $Al^{3+}$ ions and the disorder in the occupation of octahedral cation sites. Vibrational properties, which bear signatures of structure and magnetic order, are central to magneto-electric behavior of many multiferroics. In particular, Raman spectroscopy has proved to be a powerful probe to investigate magnetic-ordering induced phonon renormalization where the observed phonon anomalies below the magnetic transition temperature have been associated with the strong spin-phonon coupling [8-11].

There is no report so far of a Raman study of $AlFeO_3$. However, on a related system $GaFeO_3$ first-order Raman modes are reported [12-13] and the observed modes show anomalous temperature dependence near $T_c$ (~ 210 K) attributed to the spin-phonon interactions. In this paper, we report a detailed temperature-dependent Raman study of $AlFeO_3$ ($T_c \sim$ 250 K) with a goal to understand the phonon renormalization due to spin-phonon coupling in magnetically ordered state below $T_c$. We have also looked for phonon signatures of a ferroelectric transition at ~ 100 K arising from magnetic interactions [14]. Our study covers first-order as well as high frequency second-order and two-magnon Raman scattering. Taking inputs from first-principles density functional theory based calculations of phonons in different magnetically ordered AlFeO$_3$, our temperature dependent Raman study reveals strong phonon renormalization below $T_c$, and that its origin is in the strong spin-phonon coupling and a coupling between two-phonon modes and magnetic excitation.

## 2. METHODS

### 2.1 Experimental Details

Polycrystalline samples of AlFeO$_3$ were prepared and characterized as described in reference [15]. Unpolarised micro-Raman measurements were performed on polycrystalline pellet of AlFeO$_3$ in backscattering geometry using 514.5 nm line of an Ar-ion Laser (Coherent Innova 300) and Raman spectrometer (DILOR XY) coupled to a liquid nitrogen cooled CCD detector. Temperature variation was done from 5 K to 315 K, with a temperature accuracy of ± 0.1K using continuous flow He cryostat (Oxford Instrument).

### 2.2 Computational Details

Our first-principles calculations are based on density functional theory (DFT) with spin-density dependent exchange correlation energy functional of a generalized gradient approximated (GGA) (PerdewWang 91 (PW 91)) form [16] as implemented in the Vienna ab initio Simulation Package (VASP) [17-18]. The projector augmented wave (PAW) method [19] was used to capture interaction between ionic cores and valence electrons. An energy cutoff of 400 eV was used for the plane wave basis and integrations over the Brillouin zone of the orthorhombic crystal were sampled with a regular 4x2x2 mesh of k-points. Dyanmical matrix and phonons at $\Gamma$-point ( q = 0,0,0 ) have been obtained with a frozen-phonon method with atomic displacements of $\pm$ 0.04 Å. Numerical errors in our calculations break the symmetry of the dynamical matrix weakly and introduce an error of about $\pm$ 12

cm$^{-1}$ in the phonon frequencies.

## 3. RESULTS AND DISCUSSIONS

### 3.1. Raman Scattering from Phonons

AlFeO$_3$ has a layered structured belonging to the orthorhombic *Pna2$_1$* space group containing eight formula units i.e. 40 atoms in a unit cell, resulting in 120 normal modes, namely $\Gamma_{Fe}$ = 6A$_1$ + 6A$_2$ + 6B$_1$ + 6B$_2$, $\Gamma_{Al}$ = 6A$_1$ + 6A$_2$ + 6B$_1$ + 6B$_2$ and $\Gamma_O$ = 18A$_1$ + 18A$_2$ + 18B$_1$ + 18B$_2$ [12]. Since the inversion symmetry is lacking, Raman modes are also infrared active. There are 117 Raman modes, while A$_1$ + B$_1$ + B$_2$ are acoustic modes. Figure 1 shows the Raman spectrum at 5 K, revealing 18 modes labeled as S1 to S18 in the spectral range of 100-2200 cm$^{-1}$. Spectra are fitted with a sum of Lorentzian functions; the individual modes shown by thin lines and the resultant fit by a thick line. Our first principles density functional calculations (discussed later) suggest that the first-order Raman phonons occur below ~ 810 cm$^{-1}$. Table-I lists the experimental (at 5 K) and the calculated frequencies for disordered Anti-Ferromagnetic (AFM) state close to the experimental values are listed. Since the intensity of the mode S15 is zero above $T_c$, it is attributed to the two-magnon Raman scattering (to be discussed later). The modes S16 to S18 are assigned to second-order Raman scattering coupled with magnetic degrees of freedom.

### 3.2. Temperature Dependence of the First-Order Phonons

Figure 2 shows the mode frequencies of some of the prominent first-order phonon modes S4, S7 to S10, S13 and S14 as a function of temperature. The following observations can be made: (i) The frequencies of S4, S7, S8, S9, S10 and S13 modes show a sharp change at $T_c$. The temperature derivative of modes S4 and S10 frequencies ($\partial\omega/\partial T$) change sign at $T_c$. The frequency of mode S13 shows a jump by ~ 4 cm$^{-1}$ near $T_c$. (ii) The slope of $\omega$ with respect to temperature for the S4, S8, S9,

S10 and S14 modes show changes near 100 K. We attribute these changes to a ferroelectric transition in this system at ~ 100 K, as the pyroelectric experiments showed [14] that a polar phase exits below ~ 100 K and the reversal of polarization data with changing direction of electric field during pyroelectric current measurement demonstrates that the material is indeed a ferroelectric. The solid lines in panels are linear fits in three regions i.e. 315 K to 250 K, 250 K to 100 K and 100 K to 5 K. (iii) The temperature dependence of the mode S8 is anomalous below $T_c$ i.e frequency decreases on lowering the temperature.

The anomalies in the temperature dependence of the phonon modes S4, S7, S8, S9 and S10 near $T_c$ are similar to those in $RMnO_3$ ( R = Pr, Nd, Sm, Tb, Dy, La ), $GaFeO_3$ and $BiFeO_3$ [8, 10-13, 20-23]. The sharp change in the frequency of the mode S13 at $T_c$ can arise from subtle local structural change. Following manganites [8, 10-13, 20-23] and our theoretical calculations, the sharp changes in mode frequencies of S4, S7, S8, S9 and S10 are attributed to strong spin-phonon coupling in the magnetic phase below $T_c$.

**3.3. High Frequency Modes: Second-Order Phonon and Magnon Scattering**

Two-phonon Raman bands are related to two-phonon density of states having contribution from all the branches in the first Brillouin zone. For simplicity, we have fitted the high energy Raman band (1100 - 1800 $cm^{-1}$) with a sum of four Lorentzain (S15 to S18) where peak positions represent maxima in the two-phonon density of states. As the second-order Raman scattering involves the phonons over the entire Brillouin zone, the frequencies of the observed second-order phonons are not necessarily double of the first-order phonons at the $\Gamma$(q = 0,0,0) point. Accordingly, mode S17 can be assigned as a combination of S13 and S14 and mode S16 as overtone of mode S13 and mode S18 as overtone of mode S14.

Figure 3(a and b) shows the high frequency modes at few typical temperatures. It can be seen that the mode S15 is absent in the spectrum recorded at 265 K and above. Figure 3(d) shows the integrated intensity of the S15 mode with respect to its intensity at 250 K. Taking S13 mode as an internal marker,

Fig. 3(c) shows the intensity of S15 mode with respect to that of S13 mode. The intensity of the mode S15 is zero above $T_c$ and its intensity builds up as we lower the temperature. The vanishing of the S15 mode above 250 K suggests that it can be associated with two-magnon Raman scattering. From the energy of the two-magnon band, an estimate of the nearest neighbour exchange coupling parameter $J_o$ can be made. If spins deviations are created on the adjacent sites, the two-magnon energy is given by $J_o(2Sz-1)$ where $S$ is the spin on the magnetic site ($Fe^{3+}$ here with $S = 5/2$) and $z$ ($z = 6$) is the number of nearest neighbour to that site [24]. Using $\omega = 1240$ cm$^{-1}$ (at 5 K), the estimated value of the exchange parameter $J_o$ is ~ 5.3 meV. We note that this value is very close to our first principle calculations of $J_o$ ~ 6 meV (discussed later). As temperature is lowered below $T_c$, the S15 mode frequency decreases significantly (~ 5 %) (see Fig.4e). The frequencies of modes S16, S17 and S18 show a change in $\partial\omega/\partial T$ near the transition temperature (see Figs.4b, 4c and 4d) attributed to the possible coupling between two-phonon and magnetic excitations similar to that in other magnetic systems [8, 11, 20-30].

To ascertain the second-order nature of high frequency bands S16, S17 and S18, we plot in Fig.4(a) the sum of their intensities with respect to the intensity of the dominant first-order S13 mode. The second-order Raman intensity for a combination mode of frequency $(\omega_1 + \omega_2)$ is $[n(\omega_1)+1][n(\omega_2)+1]$, where $n(\omega)$ is the Bose–Einstein mean occupation number. The ratio of the second-order band with respect to the first-order mode of frequency $\omega_1$ will be $c[n(\omega_2)+1]$, where C is the ratio depending on the matrix elements in second-order and first-order Raman scattering. The solid line in Fig.4(a) is $1.5*[n(\omega=740 cm^{-1})+1]$ showing that the broad band (decomposed into S16, S17 and S18 modes) is due to second-order Raman scattering. We now develop theoretical understanding of our results.

### 3.4. First-principles Calculations

It is known [7] that $AlFeO_3$ has a site occupancy disorder between Fe and Al sites, with most common occurrence of anti-site disorder being between $Fe_2$ and $Al_2$ sites [7]. This disorder is taken into account by exchanging the site positions of an Fe atom at $Fe_2$ site with an Al atom at $Al_2$ site. We have also considered the anti-site disorder between $Fe_1$ and $Al_2$ sites. From the energetics, we find that the AFM state is the most stable for system with either type of anti-site disorder between Fe and Al. The AFM state with $Fe_1$-$Al_2$ anti-site disorder is higher in energy as compared to the AFM state with $Fe_2$-$Al_2$ anti-site disorder by 5.7 meV/atom, confirming the higher occurrence of $Fe_2$-$Al_2$ anti-site disorder. To facilitate a meaningful comparison with experimental Raman spectra, we simulate the structure with experimental lattice constants and relaxing internally the atomic positions using conjugate gradients algorithm.

To understand the interplay between disorder, magnetic ordering and phonons, we determine phonons at $\Gamma$-point for a chemically disordered structure with non-magnetic (NM), FM and AFM ordering (see Fig. 5). The spin-phonon coupling is analysed by examining how normal modes depend on the magnetic ordering by examining the correlation matrix between phonon eigenmodes of $AlFeO_3$ in two different magnetically ordered states. In the absence of spin-phonon coupling, the phonons would be unaffected by changes in the magnetic order and hence only the diagonal terms would be non-zero in the correlation matrix. Non-zero off-diagonal elements of the correlation matrix clearly uncover the correspondence between eigenmodes in different magnetic orders. For example, it determines which phonon modes of the AFM state relate to phonons of the FM state, giving a quantitative idea of mixing between modes due to spin-phonon coupling.

The spin-Hamiltonian has the form:

$$H = \frac{1}{2} \sum_{ij} J_{ij} \vec{S}_i . \vec{S}_j \qquad (1)$$

where, $J_{ij}$ is the exchange interaction between i[th] and j[th] ising spins $S_i$ and $S_j$. Only considering the nearest neighbour and isotropic interaction, we reduce $J_{ij}$ to $J$. The change in $J$ due to spin phonon coupling is given by second-order Taylor series expansion of $J$ w.r.t amplitude of atomic displacements $(u_{v\Gamma})$ of the $v^{th}$ $\Gamma$- phonon mode of the magnetic state [20],

$$J(u_{v\Gamma}) = J_o + \vec{u}_{v\Gamma}(\nabla_u J) + \frac{1}{2}\vec{u}_{v\Gamma}(\nabla^2_u J)\vec{u}_{v\Gamma} \quad (2)$$

Summing over all modes gives

$$H = \frac{1}{2}\sum_v \sum_{ij} [J_o + \vec{u}_{v\Gamma}(\nabla_u J) + \frac{1}{2}\vec{u}_{v\Gamma}(\nabla^2_u J)\vec{u}_{v\Gamma}].\vec{S}_i\vec{S}_j \quad (3)$$

Here, $J_o$ is the bare spin-spin coupling parameter, $\nabla_u J$ corresponds to the force exerted on the system due to change in magnetic ordering from its ground state magnetic configuration and $\nabla^2_u J$ is proportional to the change in phonon frequency $(\Delta)$ of $v^{th}$ $\Gamma$- phonon mode due to change in magnetic ordering. From the spin-Hamiltonian (Eq. 1), energies of a single pair of spins in AFM and FM states are given by, $E_{AFM} = -J_o |S|^2$ and $E_{FM} = J_o |S|^2$ respectively. The difference in the energies of AFM and FM states is directly proportional to $J_o$. The unit cell of $AlFeO_3$ used in our simulation contains 8 Fe ions where, the $i^{th}$ Fe ion is connected to $z_i$ number of other Fe ions. This gives, $J_o = (E_{FM} - E_{AFM})/(\sum_i z_i * 8 * |S|^2)$ here, $S = 5/2$ and $E_{FM} - E_{AFM} \sim 1.5$ eV from first principles calculations. Our estimate of the exchange coupling parameter $J_o$ is $\sim 6$ meV. This value is in good agreement with the one estimated from the two-magnon peak observed in Raman spectrum here. Denoting $\nabla^2_{uv} J$ as $J_2$, the change in phonon frequency $(\Delta)$ of the $\lambda^{th}$ $\Gamma$- point phonon mode is given by [20]

$$\Delta_\lambda = \frac{1}{2\mu_\lambda \omega_\lambda} \sum_v \hat{u}_{v\Gamma} J_2 \hat{u}_{v\Gamma} \tag{4}$$

Here, $\mu_\lambda$ and $\omega_\lambda$ are the reduced mass and frequency of the $\lambda^{th}$ $\Gamma$-phonon mode, respectively. We note large $\Delta$ implies stronger spin-coupling

The calculations are done for both types of disorder: $Fe_2$ at $Al_2$ site ($Fe_2$-$Al_2$) as well as $Fe_1$ at $Al_2$ site ($Fe_1$-$Al_2$). For $Fe_2$-$Al_2$ type disorder Figures 6(a) and 6(b) show the changes in $\Gamma$- point phonon frequency ($\Delta$) between FM and AFM, and NM and AFM states, respectively. The corresponding changes for $Fe_1$-$Al_2$ disorder are shown in fig. 6(c) and (d).

For connecting our results with experiment, we have listed only those calculated phonon frequencies which are close in frequency to the experimentally observed Raman active phonon modes (refer to Table-I). We assume the correlation between the experimentally observed modes which exhibit anomalies at magnetic transition and calculated spin-phonon coupling for modes with frequencies in the vicinity of the observed modes, and carry out the mode assignment. In the case of $Fe_2$-$Al_2$ anti-site disorder, Figure 6(a) and (b), corresponding to correlation between phonons of FM/NM state with AFM state, $\Delta$ and hence $J_2$ is high for mode frequencies in the neighbourhood of close to modes S1, S4 and S10 for FM-AFM coupling (see Fig 6a) and modes S4, S8, S11 and S12 (see Fig. 6b) for NM-AFM coupling. We note that $J_2$ for NM-AFM state coupling is not exactly spin-phonon coupling parameter as in the case of FM-AFM state coupling, but here it gives an estimate of the change in phonon frequencies in going from NM state to AFM ordering. In Fig.2 modes S4, S7, S8 and S10 show sharp changes in frequency at the transition temperature $T_c$ suggesting their strong coupling with spin consistent with our first-principles calculations. Another interesting observation from Fig. 6(b) is that the mode with frequency near S8 shows the largest increase in frequency in going from AFM to the NM state consistent with our experimental observation of most significant hardening of mode S8 with increase in temperature of the AFM state.

We now discuss the effect of $Fe_1$-$Al_2$ anti-site disorder, Figs. 6(c) and (d) correspond to changes in frequencies of phonons of FM and NM state correlating with those of the AFM state, respectively. With change in magnetic ordering from FM to AFM state, modes near S10 and S11 (see Fig. 6c) exhibit strong second order coupling with spin. In comparison, modes close to S4 and S5 (see Fig. 6d) show large second-order coupling for the transition from NM to AFM state. Mode S4 was observed in our Raman measurements to exhibit a significant hardening across the transition from AFM to the NM state, consistent with our calculated results.

Our first-principles analysis confirms the existence of strong spin-phonon coupling in $AlFeO_3$, and points out that the anomaly in mode S8 is primarily influences by $Fe_2$-$Al_2$ disorder, while the anomaly in S4 mode is influenced by $Fe_1$-$Al_2$ disorder additionally. Anomalous hardening of the S8 mode is due to strong spin-phonon coupling at the second-order ($J_2$) in $AlFeO_3$.

## 4. CONCLUSIONS

In conclusion, we observed a strong first-order phonon renormalization below the magnetic transition temperature of $AlFeO_3$ due to strong spin-phonon coupling. In addition, high frequency Raman bands between 1100 to 1800 cm$^{-1}$ show pronounced effects of the strong magnetic correlation below $T_c$. In particular, the intensity of mode S15 becomes zero above the transition temperature $T_c$ and hence the mode is attributed to two-magnon Raman scattering. The band position gives an estimate of spin exchange constant $J_0$ to be ~ 5.3 meV, in close agreement with the DFT calculations. With first-principles analysis we have explored the effects of magnetic ordering and (Al, Fe) disorder on phonons. Our results suggest a strong interplay between lattice and magnetic degrees of freedom which are crucial to understand the underlying physics responsible for the various exotic physical phenomena in these materials. What is equally noteworthy is that our Raman data show evidence for a phase transition to a ferroelectric phase below 100 K.

## Acknowledgments

PK, AB and SS acknowledge CSIR, India, for research fellowship. AKS acknowledge the DST, India, for financial support. UVW acknowledge DAE Outstanding Researcher Fellowship for partial financial support.

Table-I: List of the experimental observed frequencies at 5 K and calculated frequencies in $AlFeO_3$ for disordered AFM ($Fe_2$-$Al_2$ anti-site disorder) state.

| Mode Assignment | Experimental $\omega$ (cm$^{-1}$) | Calculated $\omega$ (cm$^{-1}$) |
|---|---|---|
| S1 | 156 | 154 |
| S2 | 178 | 179 |
| S3 | 198 | 197 |
| S4 | 268 | 270 |
| S5 | 328 | 331 |
| S6 | 380 | 379 |
| S7 | 425 | 425 |
| S8 | 453 | 453 |
| S9 | 498 | 499 |
| S10 | 587 | 581 |
| S11 | 650 | 654 |
| S12 | 698 | 691 |
| S13 | 738 | 733 |
| S14 | 826 | 807 |
| S15 (Two-magnon) | 1240 | |
| S16 (Overtone) | 1450 | |
| S17 (Second-order) | 1560 | |
| S18 (Overtone) | 1660 | |

**References**

[1] T. Kimura, T. Goto, H. Shintani, K. Ishizaka, T. Arima and Y. Tokura, Nature **426**, 55 (2003).

[2] W. Prellier, M. P. Singh and P. Murugavel, J. Phys. : Condens. Matter **17**, R 803 (2005).

[3] S. W. Cheong and M. Mostovoy, Nat. Mater. **6**, 13 (2007).

[4] C. N. R. Rao and C. R. Serrao, J. Mater. Chem. **17**, 4931 (2007).

[5] D. Khomski, Phyics **2**, 20 (2009).

[6] J. H. We, S. J. Kim and C. S. Kim, IEEE Tran. Mag. **42**, 2876 (1991).

[7] F. Bouree, J. L. Baudour, E. Elbadraoui, J. Musso, C. Laurent and A. Rousset, Acta Cryst. **B52**, 217 (1996).

[8] J. Laverdiere, S. Jandl, A. A. Mukhin, V. Y. Ivanov, V. G. Ivanov and M. N. Iliev, Phys. Rev. B **73**, 214301 (2006).

[9] M. O. Ramirez, A. Kumar, S. A. Denev, Y. H. Chu, J. Seidel, L. Martin, S. Y. Yang, R. C. Rai, X. Xue, J. F. Ihlefeld, N. Podraza, E. Saiz, S. Lee, J. Klug, S. W. Cheong, M. J. Bedzyk, O. auciello, D. G. Schlom, J. Orenstein, R. Ramesh, J. L. Musfeldt, A. P. Litvinchuk and V. Gopalan, Appl. Phys. Lett. **94**, 161905 (2009).

[10] P. Kumar, S. Saha, D. V. S. Muthu, J. R. Sahu, A. K. Sood and C. N. R. Rao, J. Phys. : Condens. Matter **22**, 115403 (2010).

[11] M. Viswanathan, P. S. A. Kumar, V. S. Bhadram, C. Narayana, A. K. Bera and S. M. Yusuf, J. Phys. : Condens. Matter **22**, 346006 (2010).

[12] K. Sharma, V. R. Reddy, D. Kothari, A. Gupta, A. Banerjee and V. G. Sathe, J. Phys. : Condens. Matter **22,** 146005 (2010).

[13] S. Mukherjee, R. Gupta and A. Garg, J. Phys. : Condens. Matter **23,** 445403 (2011).

[14] A. Shireen, R. Saha, S. N. Shirodkar, U. V. Waghmare, A. Sundaresan and C. N. R. Rao, arXiv:1112.5848v1.

[15] R. Saha, A. Shireen, A. K. Bera, S. N. Shirodkar , Y. Sundarayya, N. Kalarikkal, S. M. Yusuf, U. V. Waghmare, A. Sundaresan and C. N. R. Rao, J. Solid State Chem. **184**, 494 (2011).

[16] J. P. Perdew, J. A. Chevary, S. H. Vosko, K. A. Jackson, M. R. Pederson, D. J. Singh and C. Fiolhais, Phys. Rev. B **46**, 6671 (1992).

[17] G. Kresse and J. Hafner, Phys. Rev. B **47**, R558 (1993).

[18] G. Kresse and J. Furthmller, Phys. Rev. B **54**, 11 169 (1996).

[19] G. Kresse and D. Joubert, Phys. Rev. B **59**, 1758 (1999).

[20] E. Grando, A. García, J. A. Sanjurjo, C. Rettori, I. Torriani, F. Prado, R. D. Sánchez, A. Caneiro and S. B. Oseroff , Phys. Rev. B **60**, 11879 (1999).

[21] A. P. Litvinchuk, M. N. Iliev, V. N. Popov and M. M. Gospodinov, J. Phys. : Condens. Matter **16,** 809 (2004).

[22] H. Fukumara, N. Hasuike, H. Harima, K. Kisoda, K. Fukae, T. Yoshimura and N. Fujimura, J. Phys. : Condens. Matter **21,** 064218 (2009).

[23] P. Kumar, S. Saha, C. R. Serrao, A. K. Sood, and C. N. R. Rao, Pramana J. Phys. **74**, 281 (2010).

[24] M. J. Massey, U. Baier, R. Merlin and W. H. Weber, Phys. Rev. B **41**, 7822 (1990).

[25] S. J. Allen and H. J. Guggenheim, Phy. Rev. Lett. **21**, 1807 (1968).

[26] I. W. Shepherd, Phy. Rev. B **5**, 4524 (1972).

[27] M. O. Ramirez, M. Krishnamurthi, S. Denev, A. Kumar, S. Y. Yang, Y. H. Chu, E. Saiz, J. Seidel, A. P. Pyatakov, A. Bush, D. Viehland, J. Orenstein, R. Ramesh and V. Gopalan, Appl. Phys. Lett. **92**, 022511 (2008).

[28] P. A. Fleury, J. M. Worlock and H. J. Guggenheim, Phys. Rev. **185**, 738 (1969).

[29] C. H. Perry, E. Anastassakis and J. Sokoloff, Indian J. Pure and App. Phys. **9**, 930 (1971).

[30] M. N. Iliev, A. P. Litvinchuk, M. V. Abrashev, V. N. Popove, J. Cmaidalka, B. Lorenz and R. L. Meng, Phys. Rev. B **69**, 172301 (2004).

**FIGURE CAPTION**

FIG. 1. (Color online) Unpolarised Raman spectra of AlFeO$_3$ at 5 K. Solid (thin) lines are fit of individual modes and solid (thick) line shows the total fit to the experimental data.

FIG. 2. (Color online) Temperature dependence of the first-order phonon modes S4, S7, S8, S9, S10, S13 and S14. Solid lines are the linear fits in three different temperature region as described in the text.

FIG. 3. (Color online) (a and b) The temperature evolution of mode S15 at few typical temperatures. (c) Intensity ratio of S15 mode w.r.t to prominent first-order S13 mode. Solid line is the linear fit. (d) Temperature-dependence of the intensity of mode S15 w.r.t to its intensity at 250 K.

FIG 4. (Color online) (a) Intensity ratio of the high frequency band w.r.t to the prominent first-order mode. Solid line is the fitted curve as described in the text. (b,c,d and e) Temperature dependence of the mode S15, S16, S17 and S18. Solid lines are linear fits below and above $T_c$.

FIG. 5. (Color online) Distribution of phonons at $\Gamma$- point for AFM, FM and NM orderings with a Gaussian broadening of ~ 4 cm$^{-1}$.

FIG. 6. (Color online) Second order spin-phonon coupling in different magnetic states, (a, b) FM/NM-AFM states with Fe$_2$-Al$_2$ anti-site disorder respectively. And (c, d) FM/NM - AFM states with Fe$_1$-Al$_2$ anti-site disorder respectively.

FIGURE 1:

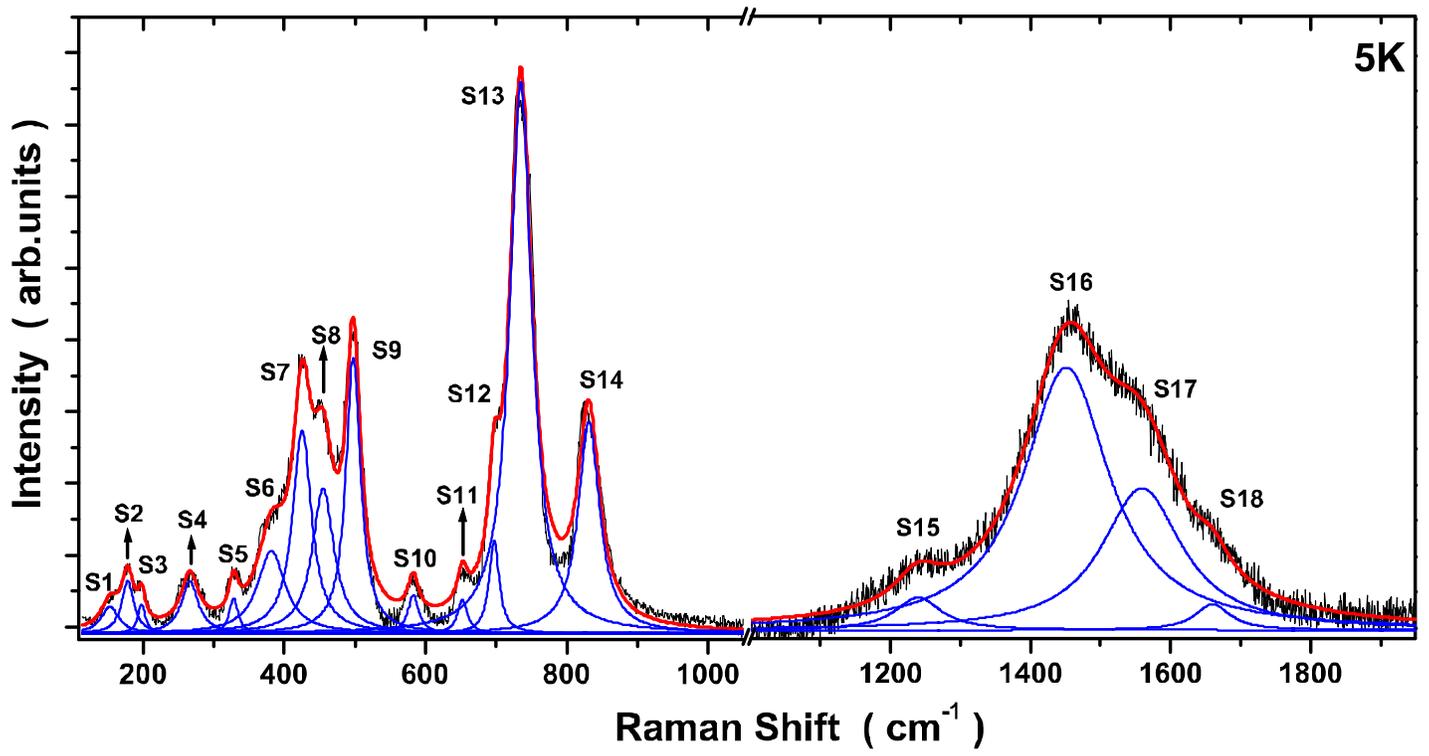

FIGURE 2:

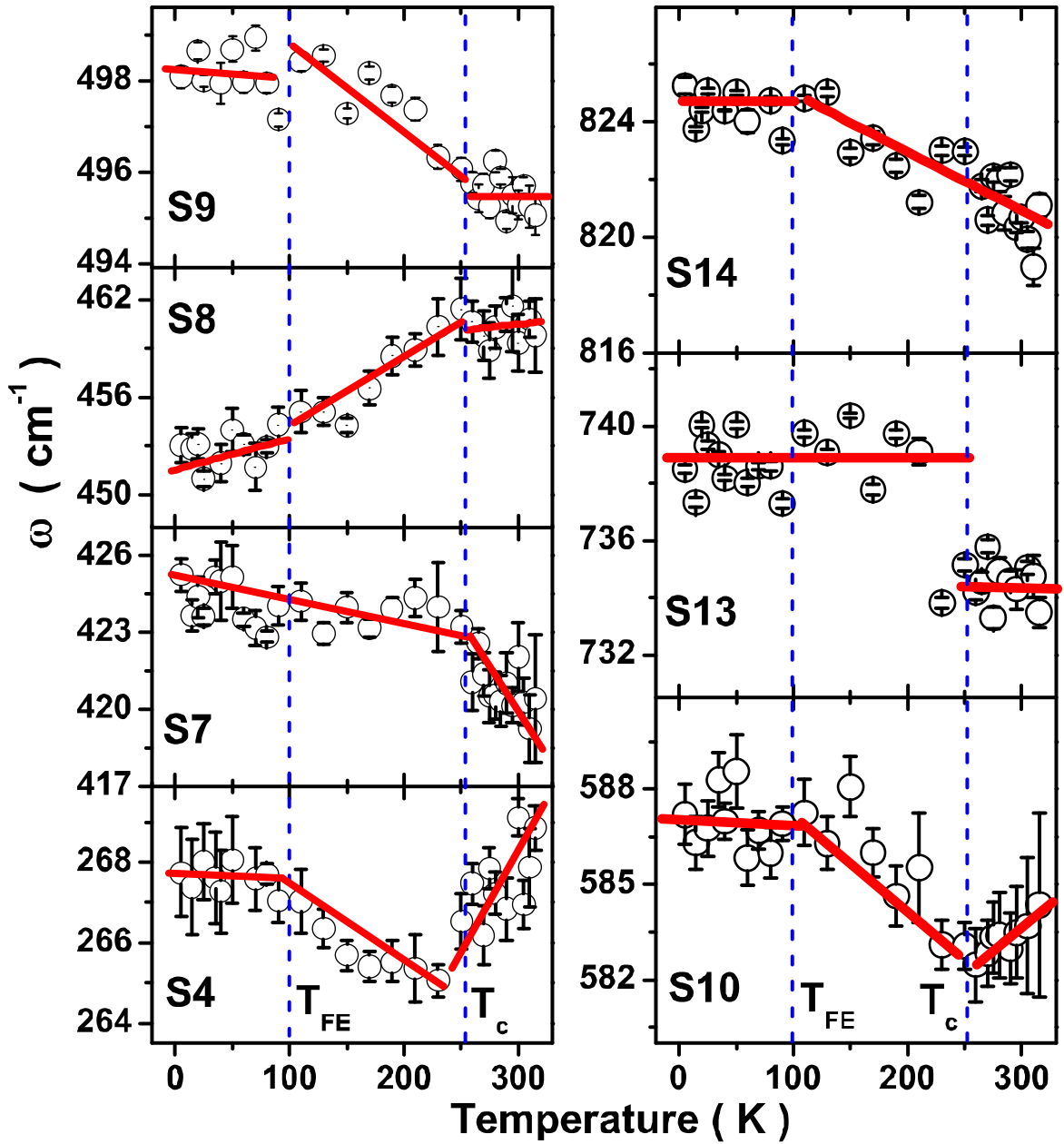

FIGURE 3:

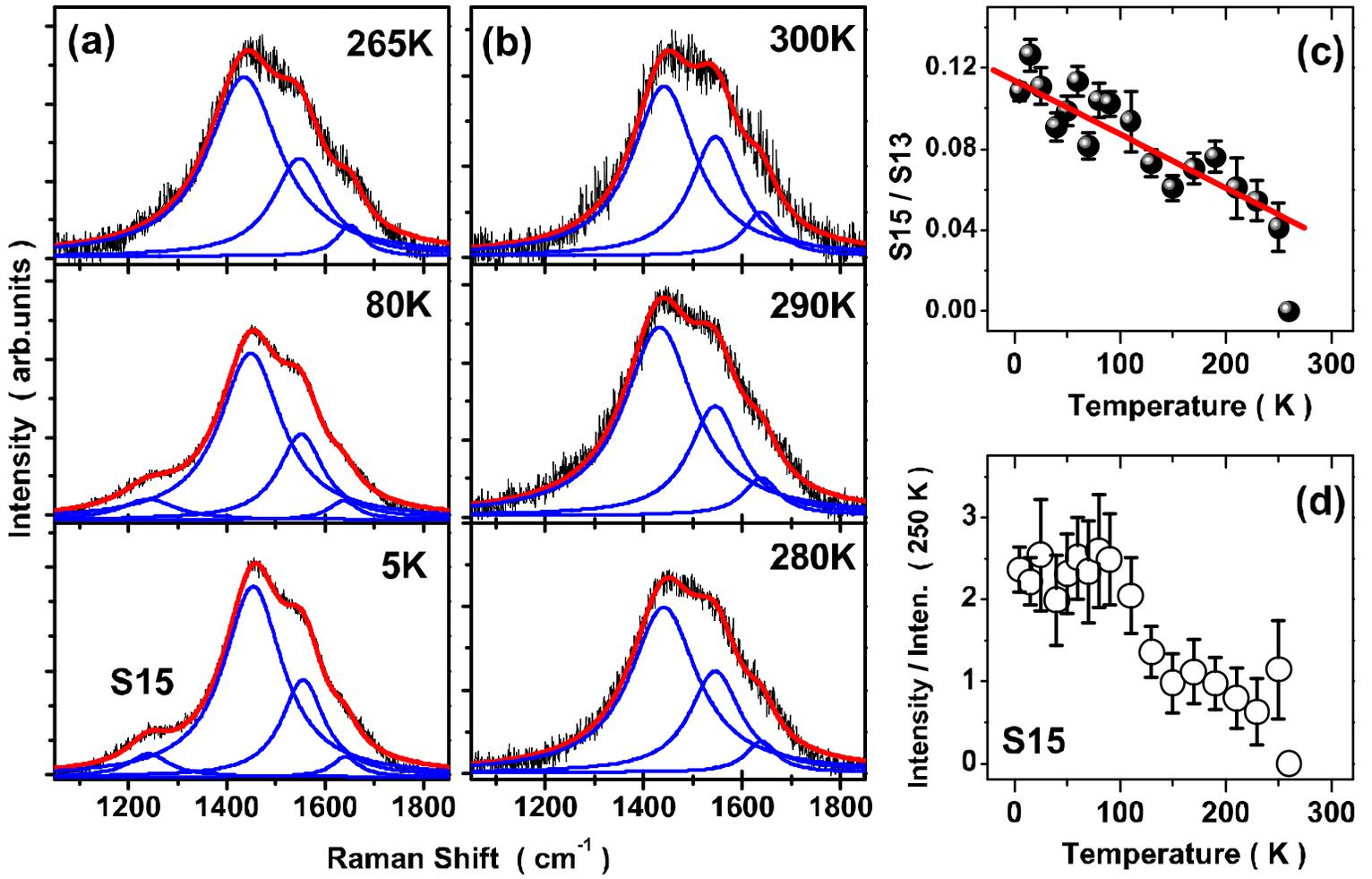

FIGURE 4:

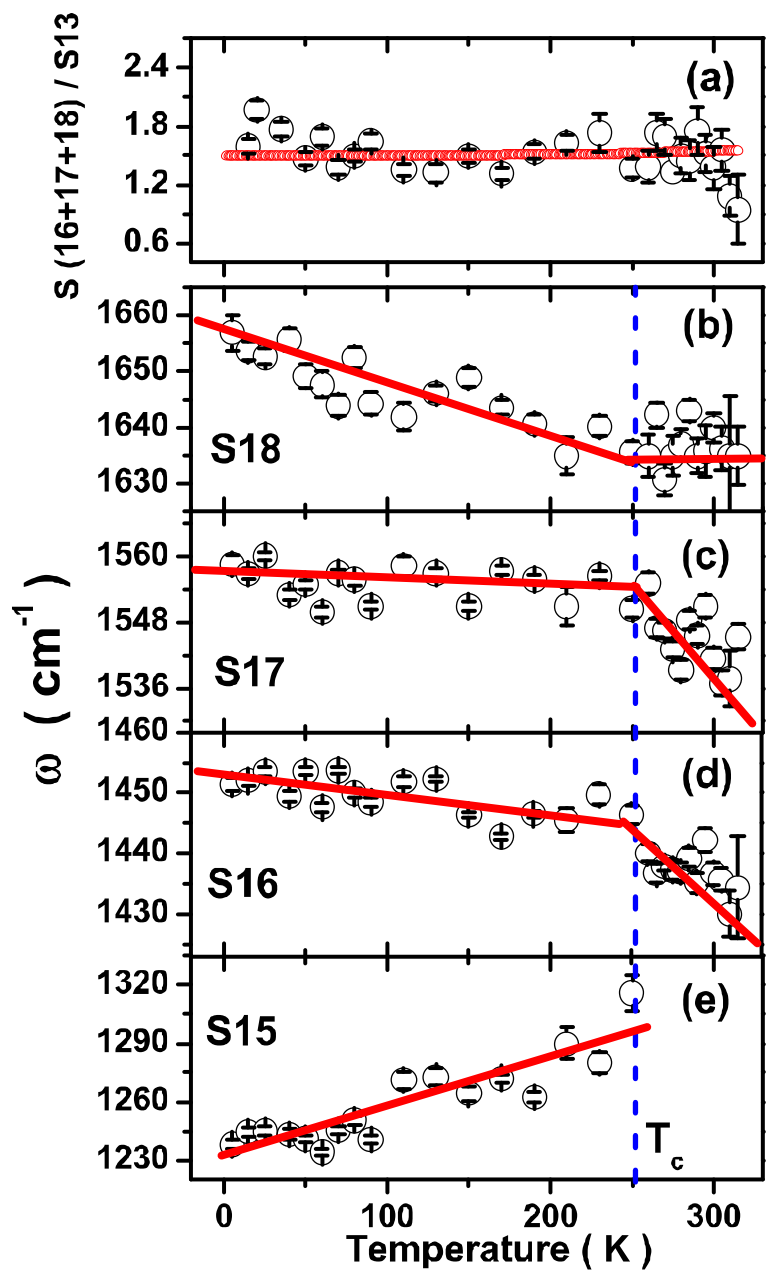

FIGURE 5:

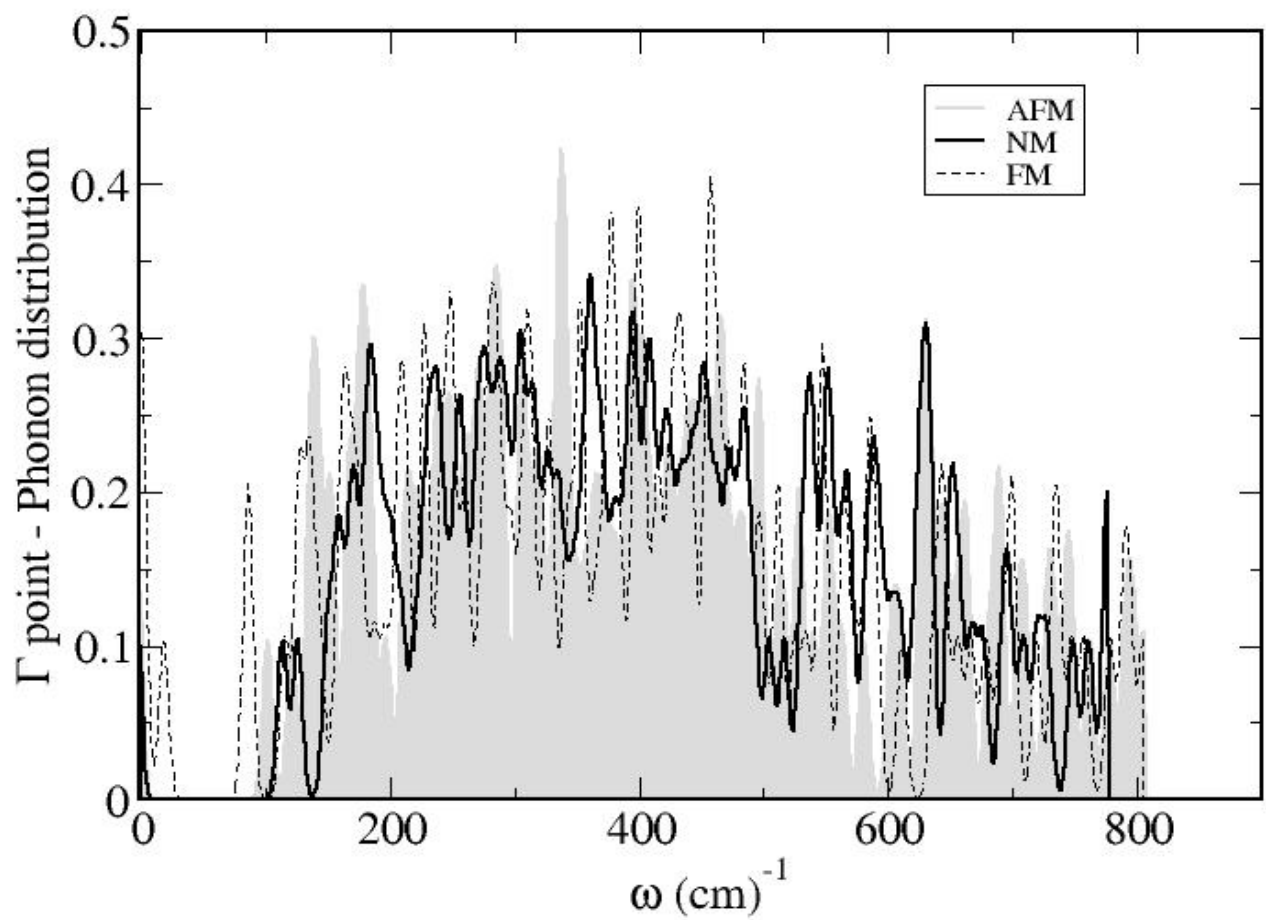

FIGURE 6:

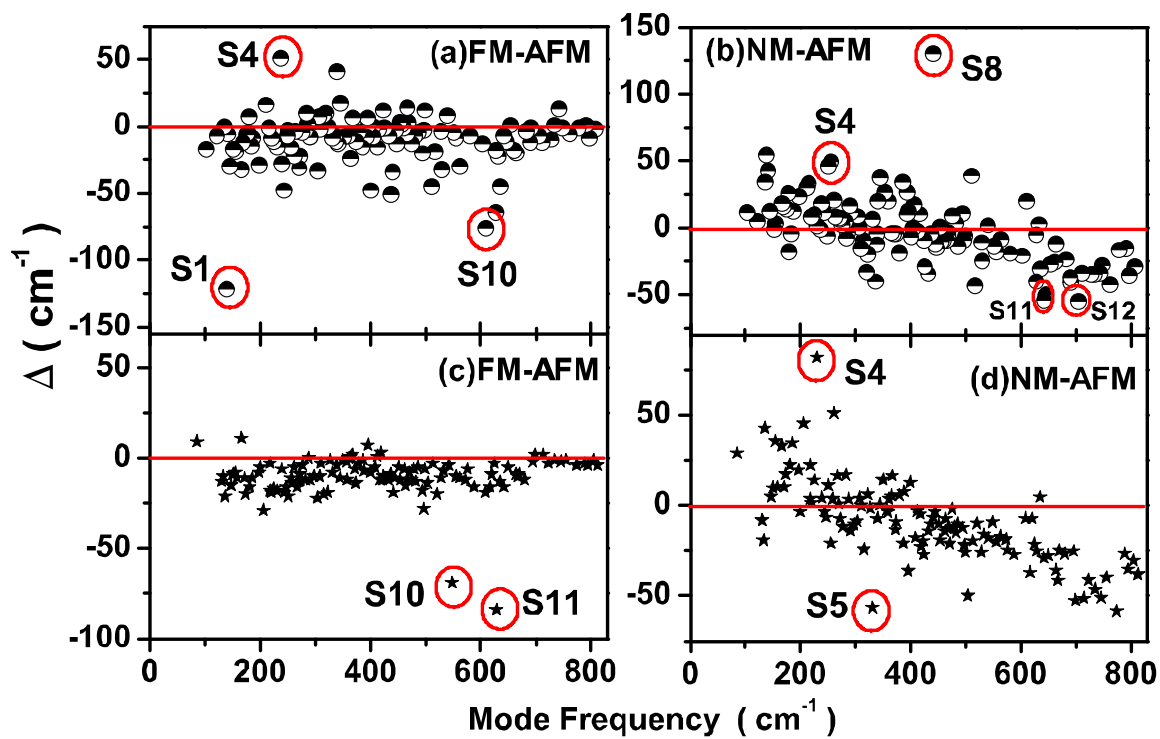